\documentclass[a4paper]{article}
\usepackage{multirow}
\usepackage[english]{babel}
\usepackage[utf8]{inputenc}
\usepackage[T1]{fontenc}
\usepackage[usenames]{color}

\usepackage[title]{appendix}
\usepackage[margin=1.5in]{geometry}

\usepackage{amsmath}
\usepackage{graphicx}
\usepackage{subcaption}
\usepackage{mwe}
\usepackage{float}
\usepackage[colorinlistoftodos]{todonotes}
\usepackage[colorlinks=true, allcolors=blue]{hyperref}
\usepackage{authblk}
\usepackage{url}
\usepackage{hyperref}
\usepackage{soul}
\usepackage{comment}
\usepackage{array,booktabs}
\usepackage[english]{babel}
\usepackage{mathrsfs}
\usepackage{blindtext}
\usepackage{tabularx}
\usepackage{stackengine}
\usepackage[normalem]{ulem}

\usepackage{tikz-cd}
\usetikzlibrary{arrows,positioning}
\usetikzlibrary{calc}

\tikzset{
  shift left/.style ={commutative diagrams/shift left={#1}},
  shift right/.style={commutative diagrams/shift right={#1}}
}
\pgfmathtruncatemacro\distance{1}

\usepackage{todonotes}
\usepackage{changebar}

\providecommand{\keywords}[1]{\textbf{Keywords ---} #1}

\title{Comparison of screening for methicillin-resistant \textit{Staphylococcus aureus} (MRSA) at hospital admission and discharge} 
\author[1]{Cole Butler}
\author[2]{Jinjin Cheng}
\author[3]{Lorena Correa}
\author[4]{María Preciado-Rivas}
\author[5]{Andrés Ríos-Gutiérrez}
\author[6]{César Montalvo}
\author[7]{Christopher Kribs}

\affil[1]{Department of Mathematics and Statistics, University of Maine, United States}
\affil[2]{College of Science, Shanghai University, China}
\affil[3]{Escuela de Ciencias Matemáticas y Tecnología Informática, Universidad Yachay Tech, Ecuador}
\affil[4]{School of Physical Sciences and Nanotechnology, Yachay Tech University, Ecuador}
\affil[5]{Department of Statistic, Universidad Nacional de Colombia, Colombia}
\affil[6]{Simon A. Levin Mathematical Computational and Modeling Sciences Center, Arizona State University, United States}
\affil[7]{Department of Mathematics, University of Texas at Arlington, United States}

\usepackage{chngpage}

\begin{document}
\maketitle
\begin{abstract}

Methicillin-resistant \textit{Staphylococcus aureus} (MRSA) is a significant contributor to the growing concern of antibiotic resistant bacteria, especially given its stubborn persistence in hospitals and other health care facility settings. In combination with this characteristic of \textit{S. aureus} (colloquially referred to as staph), MRSA presents an additional barrier to treatment and is now believed to have colonized two of every 100 people worldwide. According to the CDC, MRSA prevalence sits as high as 25-50\% in countries such as the United Kingdom and the United States.  Given the resistant nature of staph as well as its capability of evolving to compensate antibiotic treatment, controlling MRSA levels is more a matter of precautionary and defensive measures. This study examines the method of "search and isolation," which seeks to isolate MRSA positive patients in a hospital so as to decrease infection potential. Although this strategy is straightforward, the question of just whom to screen is of practical importance. We compare screening at admission to screening at discharge. To do this, we develop a mathematical model and use simulations to determine MRSA endemic levels in a hospital with either control measure implemented. We found that screening at discharge was the more effective method in controlling MRSA endemicity, but at the cost of a greater number of isolated patients. 
\end{abstract}


\section{Introduction}

Methicillin-resistant \textit{Staphylococcus aureus} (MRSA) is a bacterium  that colonizes the skin of human beings as well as their proximate environment. Although this is intrinsically true for staph, antibiotic resistance has made eradication much more difficult. The evolution of antibiotic resistance in staph, however, is not a new development.  The discovery of penicillin in the 1920s allowed for a very effective treatment for \textit{S. aureus} and other bacterial infections, remaining effective until only a few decades later when Bondi and Dietz identified the enzyme penicillinase being produced by staph - completely nullifying any power of the revolutionary antibiotic \cite{bondi1945penicillin}. Currently, more than 90\% of \textit{S. aureus} cultures are resistant to penicillin \cite{lowy2003}. Methicillin was developed as a response to penicillin resistance, but as early as the 1960s, the same decade it was developed, MRSA had already been isolated in the United Kingdom. Fifty years following initial isolation, MRSA has spread worldwide and has developed potent endemicity in health care facilities across the United States and Europe. Currently, approximately 90,000 Americans suffer from MRSA infections every year with a mortality rate of 22\% \cite{chicago}.

Since MRSA is both the most prevalent and the most destructive in hospital settings, this is the context which the following paper assumes. Screening and isolation is a very common control strategy implemented in hospitals battling MRSA outbreaks. Screening typically involves the swabbing of the nares of a patient to determine colonization, and is performed at admission. A positive result yields the placement of the patient into a region of the hospital where bacterial spread is hindered, aptly termed an isolation unit (IU). Here, further transmission of the bacteria is assumed to be zero. Preference may or may not be given to certain patients with higher susceptibility to MRSA carriage, including any individuals who have a history of hospital admission, have a history of antibiotic use, belong to a certain age group, or possess an open wound or skin infection. Recently, screening at discharge has been proposed as an alternative to screening at admission.

Several mathematical models have attempted to capture the transmission dynamics of MRSA in hospitals. Chamchod and Ruan present a compartmentalized model for MRSA that considers patients as either  uncolonized,  colonized, or infectious \cite{chamchod}.  Health care workers (HCWs) exist in their own compartments as either contaminated or uncontaminated and behave as vectors for the bacteria. Chamchod and Ruan consider MRSA transmission dynamics in light of antibiotic usage and subsequent resistance. Patients are considered at a higher risk of developing MRSA if they have a history of antibiotic usage. Cooper et al. consider additionally the contributions of the community to endemic levels in hospitals \cite{cooper}. However, the community that Cooper et al. consider is comprised entirely of previous hospital patients. The authors highlight that timing of intervention, resource provision, isolation practices, and the correct combination of procedures is the key to successful eradication. Bootsma et al. constructed two models to study MRSA transmission: one model considers transmission within a single hospital, while another model considers transmission within a system of hospitals \cite{bootsma}. In all of the aforementioned models, screening, if any, is performed at admission. 

MRSA is classified in accordance with where it originates: community-acquired MRSA (CA-MRSA) and hospital-acquired MRSA (HA-MRSA). As a result of its persistence and antibiotic resistance, MRSA is able to maintain endemic rates within health care facilities for extended periods of time. MRSA hospital endemicity yields exorbitant costs of treatment and precautions in lieu of effective antibiotic treatment. Hospitals with high endemic rates become sources of infection instead of facilities for recovery. Consequently, the attention of this research focuses on HA-MRSA only.

One aspect deserving elaboration is the notion of colonization. A patient is considered colonized when the bacteria is present on his physical person. Common places include the nares, throat, and groin \cite{kluytmans1997nasal}. Robicsek et al. estimate that MRSA colonization half-life in a patient can be up to 40 months \cite{robicsek2009duration}. Carrying the bacteria is different from being infected. Infection occurs when MRSA is allowed to enter the body, typically by way of skin lesions or wounds. Thus, from this information it can be inferred that health care workers (HCWs) are the main carriers of MRSA, as they interact with individual patients the most and are likely to be contaminated for longer periods of time due to continuous exposure to the bacteria \cite{albrich2008healthcare}. Following the example of Chamchod and Ruan, HCWs will be considered separate from the patient population and treated as vectors of the bacteria.

Screening is used to detect patients who have been colonized by MRSA. There is no unique screening procedure followed by hospitals in general. Molecular techniques, such as polymerase chain reaction (PCR) methods, are generally faster and more accurate in comparison to culture techniques. Kunori et al. estimates that the former technique is more expensive than the latter  \cite{kunori2002cost}. For the purposes of our study, we assume that the hospital uses rapid MRSA testing. The question of just how many patients should be screened is important. Universal screening-at-admission  is costlier and generally inefficient. Roth et al. found that universal screening-at-admission costs over twice as much as compared to alternative screening methods  \cite{roth2016cost}. One such common alternative is targeted screening, whereby patients deemed at high-risk of developing MRSA colonization/infection are screened. Such patients include those with frequent hospital stays, a history of antibiotic usage, or are hospitalized with skin wounds/lesions.

Identification of MRSA carriers is critical in health care settings. It is no coincidence then that optimizing how carriers are identified be of utmost importance. Using three mathematical models, each addressing a particular system (control strategies absent, screening at admission, and screening at discharge), we compare the most favored method of screening (that of admission) to screening at discharge in controlling nosocomial transmission of MRSA by using a combination of qualitative and numerical analysis methods to estimate reductions in the number of total contaminated and infected patients.

\section{Methods}

We used three mathematical models (both deterministic and stochastic) to explore different control strategies for MRSA spread in hospitals. Each model is a system of nonlinear differential equations. We developed a baseline model, which is a simple compartmental model of MRSA in a hospital absent all other control strategies. Each screening strategy is modeled similarly with corresponding changes to the baseline model. These changes are explained in the subsections to follow. 

\subsection{Baseline model}

Our model considers a town of $58,000$ with a single hospital of $600$ beds and a health care staff of $150$ HCWs \cite{chamchod}. For the baseline model, patients are considered to be uncolonized ($U$), colonized ($C$), or infected ($I$). A patient is colonized when MRSA bacteria is present on his/her body, but the bacteria has not progressed to infection. Health care workers (HCWs) are considered to be either uncontaminated ($H$) or contaminated ($H_C$). 

Admitted patients are either colonized or infected with proportions $\lambda_C$ and $\lambda_I$, respectively; they are uncolonized, otherwise. Our baseline model is represented by the following system of ordinary differential equations:
\vspace{1ex}
\begin{equation}
\begin{aligned}
\frac{dH}{dt} &= \delta H_C - \hat{\beta}_{1} H \frac{C}{N} - \hat{\beta}_{2} H \frac{I}{N} \\
\frac{dH_c}{dt} &= \hat{\beta}_{1} H \frac{C}{N} + \hat{\beta}_{2} H \frac{I}{N} - \delta H_c \\
\frac{dU}{dt} &= (1-\lambda_C - \lambda_I) \Lambda - (\mu_U + \gamma_U)U - \beta_1 U \frac{C}{N}  - \beta_2 U \frac{H_c}{N_H} - \beta_3 U \frac{I}{N} + \alpha C \\
\frac{dC}{dt} &= \lambda_C \Lambda - (\mu_C + \gamma_C) C + \beta_1 U \frac{C}{N}  + \beta_2 U \frac{H_c}{N_H} + \beta_3 U \frac{I}{N} - (\phi + \alpha) C \\
\frac{dI}{dt} &= \lambda_I \Lambda - (\mu_I + \gamma_I) I + \phi C\\
\end{aligned}
\label{eq:baselinesystem}
\end{equation}
\vspace{1ex}

\noindent
where $\beta_{1}$ denotes the transmission rate between colonized and uncolonized patients, $\beta_{2}$ refers to the transmission rate between contaminated HCWs and uncolonized patients, and $\beta_{3}$ is the transmission rate between infected and uncolonized patients. An uncolonized patient must first be colonized before becoming infected. $\mu$ and $\gamma$ are used to denote death and discharge/treatment rates of each compartment. $\phi$ is the rate at which colonized patients become infected. Colonized patients are decolonized at a rate of $\alpha$; thus $1/\alpha$ captures the average time of decolonization. $1/\delta$ gives the average time an HCW remains contaminated. $\hat{\beta}_{1}$ is the rate of contamination between uncontaminated HCWs and colonized patients, while $\hat{\beta}_2$ denotes the transmission efficiency between uncontaminated HCWs and infected patients.

The total population ($N$) is given as the sum of total HCWs ($N_H$) and total patients ($N_P$). $N_H$ is assumed constant, as well as $N_P$.  This latter assumption can be made with the correct choice of $\Lambda$, or the rate at which patients are admitted into the hospital. A patient is admitted into the hospital whenever an existing patient leaves, either by death or discharge. For the baseline model, $\Lambda = \left( \mu_{U}+\gamma_{U} \right) U+\left( \mu_{C}+\gamma_{C} \right) C+\left( \mu_{I}+\gamma_{I} \right) I$. With these assumptions, the total population within the hospital is constant.  

Patients and HCWs are assumed to mix homogeneously. Strictly speaking, the assumption of homogeneous mixing can be challenged, since most patients are confined to their rooms for the majority of their hospital stay and do not necessarily contact other patients directly. However, they may be in contact with equipment and surfaces, and thus have the potential to indirectly contaminate both HCWs and other patients. We consider these indirect contacts when calculating transmission rates.

There are two assumed mechanisms of contamination for uncontaminated health care workers. The first mechanism is contact with colonized patients while the second mechanism is contact with infected patients. We assume that a health care worker does not become contaminated from other HCWs \cite{boyce2002guideline, sopena2002staphylococcus}. Because it is possible for a HCW to become contaminated more than once in the same day, we do not account for frequency of particular patient contacts. The baseline compartmental model is shown in Figure 1.

\vspace{2ex}
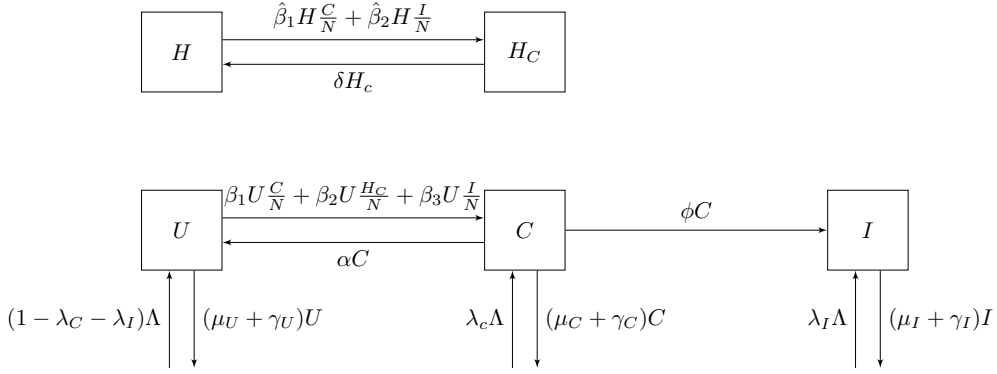
\begin{figure}[H]
	\resizebox {\textwidth}{!} {
  \begin{tikzpicture}[
  node distance=1.5cm and 4cm,
  >=latex',
  auto,
  com/.style={draw, minimum size=3.5em},
  ]
 	\node[com] (H) {$H$};
  	\node[com, right = of H] (Hc) {$H_{C}$};
	\node[com, below = of H] (U) {$U$};
    \node[com, right = of U] (C) {$C$};
    \node[com, right = of C] (I) {$I$};
    \node[below = of U, coordinate] (comu1) {};
    \node[below = of C, coordinate] (comu2) {};
    \node[below = of I, coordinate] (comu3) {};
  \path[->, shift left=1.25ex]
    (H) edge node {$\hat{\beta}_1 H \frac{C}{N} + \hat{\beta}_2 H \frac{I}{N}$}  (Hc)
    (Hc) edge node {$\delta H_c$} (H)
    (U) edge node {$\beta _{1}U\frac{C}{N} +\beta _{2}U\frac{H_{C}}{N} +\beta _{3}U\frac{I}{N}$} (C)
    (C) edge node {$\alpha C$} (U)
    (U) edge node {$( \mu _{U} +\gamma _{U}) U$} (comu1)
    (comu1) edge node {$( 1-\lambda _{C} -\lambda _{I}) \Lambda $} (U)
    (C) edge node {$( \mu _{C} +\gamma _{C}) C$} (comu2)
    (comu2) edge node {$\lambda _{c} \Lambda$} (C)
    (I) edge node {$( \mu _{I} +\gamma _{I}) I$} (comu3)
    (comu3) edge node {$\lambda _{I} \Lambda $} (I);
    \path[->]
    (C) edge node {$\phi C$} (I);

  \end{tikzpicture}
}
	\caption{Baseline model diagram.}
    \label{fig:baseline}
\end{figure}

\subsection{Screening at admission}

For the model for screening at admission, we maintain the structure of the baseline model with the addition of the isolation compartment, denoted by $Z$, henceforth referred to as the isolation unit (IU). For simplicity, we assume that the IU has infinite capacity. If a patient is tested positive for MRSA at admission, he/she will be moved to the IU for the remainder of his/her time in the hospital. Successful screening occurs at proportion $\rho$. No distinction is made between infected and colonized patients when screened for MRSA. Only newly admitted patients may be placed in the IU, with the exception being an identified infected patient in the hospital according to some rate $\kappa$. $1/\kappa$ is taken to be the sum of the average incubation period of MRSA infection ($4.5$ days) and the average duration of culture and sensitivity testing ($2.5$ days according to \cite{hal2007}). The model representing the aforementioned MRSA hospital dynamics is as follows:\\
\vspace{2ex}
\begin{equation}
\begin{aligned}
\frac{dH}{dt} &= \delta H_C - \hat{\beta}_1 H \frac{C}{N} - \hat{\beta}_2 H \frac{I}{N}\\
\frac{dH_c}{dt} &= \hat{\beta}_1 H \frac{C}{N} + \hat{\beta}_2 H \frac{I}{N} - \delta H_c\\
\frac{dU}{dt} &= (1-\lambda_C - \lambda_I) \Lambda - (\mu_U + \gamma_U)U - \beta_1 U \frac{C}{N}  - \beta_2 U \frac{H_c}{N} - \beta_3 U \frac{I}{N} + \alpha C\\
\frac{dC}{dt} &= \lambda_C \Lambda(1-\rho) - (\mu_C + \gamma_C) C + \beta_1 U \frac{C}{N}  + \beta_2 U \frac{H_c}{N} + \beta_3 U \frac{I}{N} - (\phi + \alpha) C\\
\frac{dI}{dt} &= \lambda_I \Lambda (1-\rho) - (\mu_I + \kappa) I + \phi C\\
\frac{dZ}{dt} &= (\lambda_C + \lambda_I) \Lambda \rho  + \kappa I - (\mu_Z +\gamma_Z)Z\\
 \end{aligned}
 \label{eq:admission}
 \end{equation}\\
 \vspace{1ex}
 
\noindent
For this model, $\displaystyle \Lambda=\displaystyle \left( \mu_{U}+\gamma_{U} \right) U+\left( \mu_{C}+\gamma_{C} \right) C+\left( \mu_{I} \right) I+\left( \mu_{Z}+\gamma_{Z} \right) Z$. Note that patients infected with MRSA are not discharged, but are treated in isolation. As with the baseline model, the population remains constant. Note also that we omit consideration of $Z$ regarding transmission between contaminated and uncontaminated groups. This is because we assume that $Z\ll N$. Admitted patients tested positive for MRSA move into the IU at a rate given by $(\lambda_C + \lambda_I)\rho \Lambda$. Patients in isolation are assumed to die at a rate of $\mu_Z$ and are discharged/treated at a rate of $\gamma_Z$. Patients infected with MRSA are not treated outside the IU. The schematic for this system is given in Figure 2.\\
\vspace{-3ex}

\begin{figure}[H]
	\resizebox {\textwidth}{!} {
  \begin{tikzpicture}[
  node distance=1.5cm and 4.5cm,
  >=latex',
  auto,
  com/.style={draw, minimum size=3.5em},
  ]
 	\node[com] (H) {$H$};
  	\node[com, right = of H] (Hc) {$H_{C}$};
	\node[com, below = of H] (U) {$U$};
    \node[com, right = of U] (C) {$C$};
    \node[com, node distance=3.5cm, right = of C] (I) {$I$};
    \node[com, node distance=3.5cm, right = of I] (Z) {$Z$};
    \node[below = of U, coordinate] (comu1) {};
    \node[below = of C, coordinate] (comu2) {};
    \node[below = of I, coordinate] (comu3) {};
    \node[below = of Z, coordinate] (comu4) {};
  \path[->, shift left=1.25ex]
    (H) edge node {$\hat{\beta}_1 H \frac{C}{N} + \hat{\beta}_2 H \frac{I}{N}$}  (Hc)
    (Hc) edge node {$\delta H_c$} (H)
    (U) edge node {$\beta _{1}U\frac{C}{N} +\beta _{2}U\frac{H_{C}}{N} +\beta _{3}U\frac{I}{N}$} (C)
    (C) edge node {$\alpha C$} (U)
    (U) edge node {$( \mu _{U} +\gamma _{U}) U$} (comu1)
    (comu1) edge node {$( 1-\lambda _{C} -\lambda _{I}) \Lambda $} (U)
    (C) edge node {$( \mu _{C} +\gamma _{C}) C$} (comu2)
    (comu2) edge node {$\lambda _{c} \Lambda (1-\rho)$} (C)
    (I) edge node {$ \mu _{I}  I$} (comu3)
    (comu3) edge node {$\lambda _{I} \Lambda (1-\rho)$} (I)
    (Z) edge node {$( \mu _{Z} +\gamma _{Z}) Z$} (comu4)
    (comu4) edge node {$( \lambda _{C} +\lambda _{I}) \Lambda \rho$} (Z);
    
    \path[->]
    	(C) edge node {$\phi C$} (I)
		(I) edge node {$\kappa I$} (Z);
  \end{tikzpicture}

}
	\caption{Screening at admission model diagram.}
    \label{fig:admission}
\end{figure}
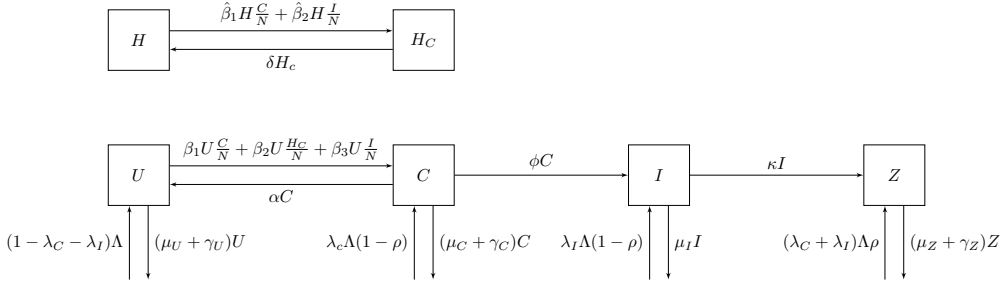

\subsection{Screening at discharge}


In order to reformulate the model so as to consider screening at discharge, we compartmentalize the community in terms of flagged ($F$) and unflagged ($F_U$) individuals. Patients are flagged if they test positive for MRSA at discharge, and are unflagged otherwise. Patients who are flagged, when readmitted to the hospital, are placed in the isolated compartment. Our model then  becomes:
\vspace{1ex}
\begin{equation}
\begin{aligned}
\frac{dH}{dt} &= \delta H_C - \hat{\beta}_1 H \frac{C}{N} - \hat{\beta}_2 H \frac{I}{N}\\
\frac{dH_C}{dt} &= \hat{\beta}_1 H \frac{C}{N} + \hat{\beta}_2 H \frac{I}{N} - \delta H_c\\
\frac{dU}{dt} &= (1-\lambda_C - \lambda_I) \Lambda \Big(\frac{F_U}{kF + F_U}\Big) - (\mu_U + \gamma_U)U - \beta_1 U \frac{C}{N}  - \beta_2 U \frac{H_C}{N} \\&\quad- \beta_3 U \frac{I}{N} + \alpha C\\
\frac{dC}{dt} &= \lambda_C \Lambda \left(\frac{F_U}{kF + F_U}\right) - (\mu_C + \gamma_C) C + \beta_1 U \frac{C}{N}  + \beta_2 U \frac{H_C}{N} + \beta_3 U \frac{I}{N} \\&\quad- (\phi + \alpha) C\\
\frac{dI}{dt} &= \lambda_I \Lambda \left(\frac{F_U}{kF + F_U}\right)- (\mu_I+\kappa) I + \phi C\\
\frac{dZ}{dt} &= \Lambda \Big(\frac{kF}{kF + F_U}\Big)+\kappa I - (\mu_Z + \gamma_Z)Z\\ 
\frac{dF}{dt} &= \rho(\gamma_C C + (1-\tau)\gamma_Z Z) - \Lambda \Big(\frac{kF}{kF + F_U}\Big) - \mu_{F}F\\
\frac{dF_U}{dt} &= (1-\rho)(\gamma_C C + (1-\tau)\gamma_Z Z) + \gamma_U U + \tau \gamma_Z Z- \Lambda \left(\frac{F_U}{kF + F_U}\right)
\\&\quad-\mu_{F_U}{F_{U}} + b_{F_U}\\
\end{aligned}
\label{eq:discharge}
\end{equation}\\ 
\noindent
In addition to the previous model, success of patient treatment is included. Proportion $\tau$ of treatments are successful of complete eradication and fail otherwise. We also consider the factor $k$, which represents the number of times more likely that a flagged patient is to be readmitted to the hospital as compared to an unflagged patient. Consequently, the total admission into the hospital is given by $\Lambda = \left( \mu_{U}+\gamma_{U} \right) U+ \left( \mu_{I} \right) I+\left( \mu_{C}+\gamma_{C} \right) C+\left( \mu_{Z}+\gamma_{Z} \right) Z$ in order to retain a constant hospital population. The unflagged population is comprised of the wider community as well as patients who were not identified as MRSA-positive when they were discharged from the hospital. Recruitment rate and death rate for the unflagged group are denoted by $b_{F_U}$ and $\mu_{F_U}$, respectively. Individuals in the flagged compartment die at a rate of $\mu_F$. The birth and death rates of the community were chosen so that the community population is asymptotically constant. The disease dynamics of this model is represented graphically in Figure 3.\\
\vspace{-2ex}
\begin{figure}[H]
\centering
\includegraphics[width=1\textwidth]{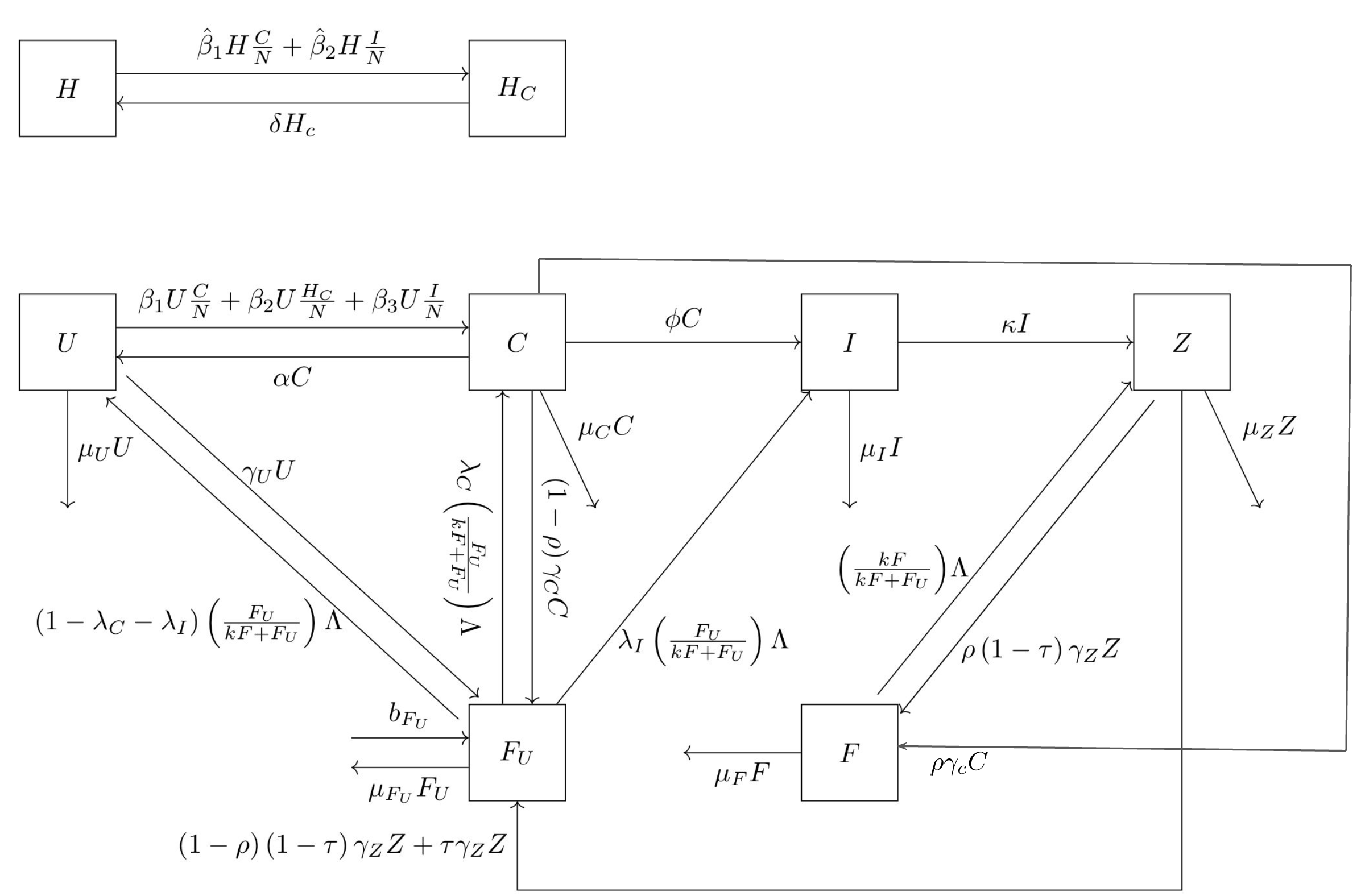}
\caption{Screening at discharge model diagram.}
\label{fig:discharge}
\end{figure}

\subsection{Parameter estimation}

All model parameters are defined, and estimates given, in Tables 1 and 2. Several parameters discussed prior deserve further elaboration, contained within this subsection. For $\beta_2$, the transmission rate between contaminated HCWs and uncolonized patients, we assumed that patients could not be colonized more than once during a single day, and that patients only make (direct) contact with HCWs (note that this is not contradictory, as patients can contact other patients via indirect means, such as shared surfaces). Reference \cite{grundmann2002risk} reports that HCWs make 7.6 contacts per patient per day. The proportion of successful transmission is taken to be 0.01 \cite{grundmann2002risk}. With this, the rate of successful transmission per day is $1-(1-0.01)^{7.6}=0.0735$.

If HCWs make 84 patient contacts per day \cite{grundmann2002risk}, then the rate of transmission from colonized patients to HCWs is $\hat{\beta}_1 = 84 \times 0.152$, where 0.152 is the probability of successful contamination. For convenience, we assume that the probability of contamination is double that for infected patients. However, since infected patients are only considered to be truly infectious for 7 days (before they move to isolation, in the case of the models with control strategies), the true rate is given as $\hat{\beta}_2 = 0.304\times 7/16 \times 84 = 11.17$. For the baseline model, which lacks an isolation compartment, corresponding adjustments would have to be made when computing these parameters.

$1/\delta$ gives the average time an HCW remains contaminated. Because data for this term is either lacking or varies greatly (e.g. an HCW can become decontaminated by merely washing his hands or an HCW can be colonized with MRSA for weeks at a time), we computed $\delta$ numerically based on the findings of Albrich and Harbarth (2008), who found that average MRSA carriage amongst HCWs is around $4.6\%$ \cite{albrich2008healthcare}. The value of $\delta$ varies between models in accordance with this figure. The models with control strategies, for example, will have lower values of $\delta$ than the baseline model as the existence of an IU restricts contamination of HCWs. For the model with screening at admission, $\delta = 48.23$ day$^{-1}$, while for the model with screening at discharge, $\delta = 50$ day$^{-1}$. For the baseline model, $\delta$ was determined to be $\delta = 108$ day$^{-1}$. The reason for the value of $\delta$ being large for the baseline model is due to the fact that the model lacks any means of patient isolation in the event of colonization/infection, and thus infected patients are infectious for the entire duration of their stay in the hospital.

$\gamma_Z$ and $\mu_Z$ are assumed to be averages of the discharge/treatment and death rates, respectively, of colonized and infected patients. That is, $\displaystyle \gamma_Z = \frac{\gamma_I + \gamma_C}{2}$ and $\displaystyle \mu_Z = \frac{\mu_I + \mu_C}{2}$. The term $\kappa$ represents the rate at which patients develop MRSA infection while in the hospital, are identified as having MRSA, and are subsequently isolated. Assuming a 4.5-day incubation period, followed by a 2.5-day period for culture and susceptibility testing, our value of $\kappa$ comes out to $0.13$. This value is close to the value of $0.14$ used by $\cite{bootsma}$. Since patients who develop infection are identified and isolated over the span of 7 days and are assumed to stay in the hospital for 16 days, each transmission rate concerning infected patients is multiplied by a factor of $7/16$, as they are assumed to be no longer infectious in isolation. Note that this only applies to the models where a control strategy is present.

$k$ represents the  number of times more likely that a flagged patient is to be readmitted to the hospital as compared to an unflagged patient. Stochastic simulations revealed that regardless of our value of $k$, the rate of patient admission from the flagged compartment would approach a stable equilibrium. This follows intuitively from the fact that, for large $k$, $F$ will become small quickly and remain small for $t\rightarrow\infty$. On the other hand, if $k$ is small, $F$ will remain large and remain so for all $t$. The death rate for either community compartment is just the average lifespan of an individual in the United States, and the birth rate of the unflagged compartment is chosen so that the population of the community is asymptotically constant.

\def\myarm{1cm}
\def\myangle{0}
\tikzset{
  arm/.default=1cm,
  arm/.code={\def\myarm{#1}}, 
  angle/.default=0,
  angle/.code={\def\myangle{#1}} 
}

\tikzset{
    myncbar/.style = {to path={
        let
            \p1=($(\tikztotarget)+(\myangle:\myarm)$)
        in
            -- ++(\myangle:\myarm) coordinate (tmp)
            -- ($(\tikztotarget)!(tmp)!(\p1)$)
            -- (\tikztotarget)\tikztonodes
    }}
}

\begin{table}[ht]
	\begin{adjustwidth}{-.5in}{-.5in}  
    \caption{Parameter definitions, values, and references}
	\begin{center}
		\begin{tabular}{l c c c}
			\textbf{Parameter definition} & 
			\textbf{Parameter} & \textbf{Values} & 					\textbf{Reference}\\
			\midrule
			Total number of patients & $N_{P}$ & 600  & N/A \\
			\midrule
			Total number of HCWs & $N_{H}$ & 150 & N/A \\
			\midrule
			Colonized proportion of newly admitted patients & $\lambda_{C}$ & $0.0374$ & \cite{fishbain2003nosocomial}  \\
			\midrule
			Infected proportion of newly admitted patients & $\lambda_{I}$ & 0.0067 & \cite{seybold2006emergence}\\
			\midrule
			Death rate of uncolonized patients & $\mu_U$ & $5.58$x$10^{-5}$ day$^{-1}$ & \cite{hall2013trends}\\
			\midrule
			Death rate of colonized patients & $\mu_{C}$ & $8.25$x$10^{-5}$ day$^{-1}$ & 			\cite{mendy2016staphylococcus} \\
			\midrule
			Death rate of infected patients & $\mu_{I}$ & $4.87$x$10^{-4}$ day$^{-1}$ & \cite{klevens2007invasive} \\
			\midrule
			Death rate of isolated patients & $\mu_{Z}$ & $2.85$x$10^{-4}$ day$^{-1}$ & estimated\\
			\midrule
			Death rate of unflagged individuals & $\mu_{F_U}$ & $3.48$x$10^{-5}$ day$^{-1}$ & N/A \\
			\midrule
			Death rate of flagged individuals& $\mu_{F}$ & $3.48$x$10^{-5}$ day$^{-1}$ & N/A \\
			\midrule
			Birth rate of community & $b_{F_U}$ & $2.018$ day$^{-1}$ & N/A \\
			\midrule
			Discharge rate of uncolonized patients & $\gamma_U$ & $0.189$ day$^{-1}$ & \cite{fishbain2003nosocomial} \\
			\midrule
			Discharge rate of colonized patients & $\gamma_C$  & $0.143$ day$^{-1}$ & \cite{davis2004methicillin} \\
			\midrule
			Treatment rate of infected patients & $\gamma_I$ & $0.063$ day$^{-1}$ & \cite{dagata2005}, \cite{Hassoun2017}, \cite{cosgrove2005} \\
			\midrule
			Treatment rate of isolated patients & $\gamma_Z$ & $0.1015$ day$^{-1}$ & estimated \\
			\midrule
			Decontamination rate of HCWs & $\delta$ & varies  & \cite{grundmann2002risk} \\
			\midrule
			Decolonization rate of colonized patients & $\alpha$ & $0.001$ day$^{-1}$ & \cite{mendy2016staphylococcus}, \cite{chamchod}\\
			\midrule
			Rate of progression from colonized to infected & $\phi$ & $0.04$ day$^{-1}$ & \cite{chamchod} \\
			\midrule
			Rate of progression from infected to isolated & $\kappa$ & $0.13$ day$^{-1}$ & estimated \\
			\midrule
			Proportion of successful treatment & $\tau$ & $0.68$ & \cite{mollema2010}\\
			\midrule
			Screening proportion & $\rho$ & varies & N/A \\
			\midrule
			\end{tabular}
		\label{tab:parameters}
		\end{center}
	\end{adjustwidth}
\end{table}


\section{Analysis}

For each model we calculate an adjusted reproduction number using the next generation matrix method \cite{diekmann1990nextgenmat}. We then perform a sensitivity analysis to determine the extent to which each parameter affects this value, and conclude by looking at the endemic equilibria of our models. 

\subsection{Reproduction Number Calculation}

\subsubsection{Baseline model}
In either model, the HCW population $N_H$ is kept constant. Because of this constraint, that can be expressed as $H = N_H - H_C$, only one from the two equations related to HCWs is needed. For this analysis we used the equation describing $H_C$ population, that is, the second equation in \eqref{eq:baselinesystem}.
To further simplify, the patient population $N_P$ in the hospital is also kept constant. We do this by setting the rate of patients being admitted per unit time, $\Lambda$, to be a function of the number of patients leaving the hospital,
\begin{equation}
\Lambda = \omega_{U}U + \omega_{C} C +\omega_{I}I,
\end{equation}
where $\omega_{J}$ is the sum of the death ($\mu_J$) and discharge ($\gamma_J$) rates of compartment $J$, i.e., $\omega_{J} = \gamma_{J} + \mu_{J}$. This leads to a second constraint, $U = N_P - C - I$, that also lets us use an equation less in \eqref{eq:baselinesystem}.

The disease-free equilibrium (DFE) is obtained when the contaminated and infected populations are zero, i.e. when we set
\begin{equation}
\begin{aligned}
H^*_{C}={C}^{*}={I}^{*}&=0,\\
{H}^{*}&=N_{H},\\
{U}^{*}&=N_{P}.\\
\end{aligned}
\label{eq:dfe_baseline}
\end{equation}
For the system described by \eqref{eq:baselinesystem}, a DFE does not exist when either $\lambda_C>0$ or $\lambda_I>0$. If this were the case, then colonized and infected patients would be admitted at each time step, forbidding the existence of a hospital state absent any contaminated patients. Nonetheless, these parameters can be set to zero to allow insight into the spread of MRSA bacteria within the hospital. That is, the model is simplified to consider the case where all newly admitted patients are uncolonized ($\lambda_I = \lambda_C = 0$). By doing so we were able to calculate an \textit{adjusted} reproduction number, denoted by $R_0$, for the baseline model as explained hereunder.

We employed the next-generation matrix method \cite{diekmann1990nextgenmat,VANDENDRIESSCHE2002nextgenmat}, in which the basic reproduction number is the largest eigenvalue, or spectral radius, of $F{V}^{-1}$, where $F$ and $V$ are the Jacobian matrices evaluated at the DFE of vectors $\mathscr{F}$ and $\mathscr{V}$.
Specifically, $\mathscr{F}$ is a vector whose entries are terms that account for newly contaminated patients and contaminated HCWs. Newly sick individuals enter either the contaminated HCW compartment, $H_{C}$, or the colonized patient compartment, $C$. Whereas, $\mathscr{V}$ contains terms corresponding to transitions and outflow of patients and HCWs from these compartments. Thus, using the next-generation matrix method, we find that

\begin{table}[!t]
	\begin{adjustwidth}{-.5in}{-.5in} 
	\caption{Transmission rates, values, and references for the models with screening}

\begin{center}
\begin{tabular}{l c c c}
\textbf{Parameter definition} & \begin{tabular}{l}
\textbf{Parameter} \\ 
\textbf{symbol}%
\end{tabular}& \begin{tabular}{l}
\textbf{Parameter} \\ 
\textbf{values}%
\end{tabular} & \textbf{Reference}\\
\midrule
Rate of patient colonization after contact \\\quad w/colonized patients &$\beta_1$ & $0.27$ day$^{-1}$ & \cite{grundmann2002risk}, \cite{d2009modeling} \\
\midrule
Rate of patient colonization after contact \\\quad w/contaminated HCWs & $\beta_2$ & $0.0735$ day$^{-1}$ & \cite{grundmann2002risk}  \\
\midrule
Rate of patient colonization after contact \\\quad w/infected patients & $\beta_3$ & $0.03$ day$^{-1}$ & \cite{dagata2005} \\
\midrule
Rate of HCW contamination after contact \\\quad w/colonized patients & $\hat{\beta}_1$ & $12.77$ day$^{-1}$ & \cite{grundmann2002risk}, \cite{spetz2008} \\
\midrule
Rate of HCW contamination after contact \\\quad w/infected patients & $\hat{\beta}_2$ & $11.17$ day$^{-1}$ & estimated \\
\midrule
\end{tabular}
\label{tab:parameters1}
\end{center}
\end{adjustwidth}

\end{table}

\begin{equation*}
\mathscr{F} = 
	\begin{pmatrix} 
    	\hat{\beta}_1 \frac{C(N_H-H_C)}{N} + \hat{\beta}_2 \frac{I(N_H-H_C)}{N} \\[6pt] 
		\beta_1 \frac{C(N_P-C-I)}{N} + \beta_2 \frac{H_C(N_P-C-I)}{N} + 			\beta_3 \frac{I(N_P-C-I)}{N} \\[6pt] 
        0 \\[6pt] 
    \end{pmatrix} 
\end{equation*} and
\begin{equation*}
\mathscr{V} = 
	\begin{pmatrix} 
    	\delta H_c\\[6pt] 
		(\alpha + \phi + \omega_C)C\\[6pt]
        \omega_I I - \phi C \\[6pt]
    \end{pmatrix}.
\end{equation*}

\noindent
The $F$ and $V$ matrices are then
\begin{equation*}
F=\begin{pmatrix} 0 & \frac {\hat{\beta}_1 N_H}{N}  & \frac {\hat{\beta}_2 N_H}{N}  \\[6pt] 
\frac { \beta_2 N_P }{N}  & \frac {\beta_1 N_P }{N}  & \frac { \beta_3 N_P  }{N}  \\[6pt]
0 & 0 & 0 \end{pmatrix}\\ \quad \text{and} \quad V=\begin{pmatrix} \delta  & 0 & 0 \\[6pt] 0 & \alpha +\phi +{ \omega  }_{ C } & 0 \\[6pt] 0 & -\phi  & { \omega  }_{ I } \end{pmatrix}
\end{equation*}

Each element $n_{ij}$ of the next-generation matrix is the average number of new colonized or infected individuals of the $i$th compartment produced by the interaction with or progression from individuals of the $j$th compartment, at each time step. For example, the first element is zero because we assumed that HCWs could not contaminate each other. Proceeding with our calculations,
\begin{equation}
 F{ V }^{ -1 }=\begin{pmatrix} 0 & \frac { N_H^* \left(\hat{\beta}_2\phi + \hat{\beta}_1 \omega_I\right)}{\left(\alpha +\phi + \omega_C\right)\omega_I}  & \frac { N_H^*\hat{\beta}_2}{\omega_I}  \\[6pt] \frac { N_P^*\beta_2}{ \delta  }  & \frac {N_P^*\left( \beta_3\phi +\beta_1\omega_I\right)}{\left(\alpha +\phi + \omega_C\right)\omega_I}  & \frac { N_P^*\beta_3}{ \omega_I }  \\[6pt] 0 & 0 & 0 \end{pmatrix}.
\end{equation}
where $ {N_P}^* = N_P/N$ and ${N_H}^* = N_H/N$. The reproduction number of the adjusted system can thus be calculated as:
\begin{equation}
\begin{split}
R_0 = &\frac{1}{2}\frac{N_P^*(\beta_3 \phi + \beta_1 \omega_I)}{(\alpha + \phi +\omega_C) \omega_I} +\\  &
\frac{1}{2}\sqrt{\left(\frac{N_P^*(\beta_3 \phi + \beta_1 \omega_I)}{(\alpha+\phi+\omega_C)\omega_I}\right)^2 + 4\left(\frac{N_H^* (\hat{\beta}_2 \phi + \hat{\beta}_1 \omega_I)}{(\alpha+\phi + \omega_C) \omega_I}\right) \left(\frac{N_P^* \beta_2}{ \delta}\right)}.
\label{eq:r0_baseline}
\end{split}
\end{equation}
Equivalently, we can represent the reproduction number as
\begin{equation}
R_0 = \frac{1}{2}\left( R_P + \sqrt{R_P^2 + 4\cdot R_H^2}\right)
\label{eq:ro_form}
\end{equation}
where $R_P$ is the colonization/infection potential of patients and $R_H$ is the contamination potential of HCWs. These two values represent processes occurring simultaneously: a direct transmission between patients and a two-step cycle of transmission between patients and HCWs. 

$R_H$ is the geometric mean of (1) the average number of new HCWs contaminated per colonized/infected patient and (2) the average number of new patients colonized per contaminated HCW. The first of these factors has two terms, each accounting for different transmission pathways: one direct (HCWs being contaminated by colonized patients) and other indirect (HCWs being contaminated by infected patients). This term is
\begin{equation}
{R_H}=\sqrt{ N_H^*\left( \frac{\hat{\beta}_1}{\alpha +\phi +\omega_C} +\frac {\phi}{\alpha +\phi +\omega_C} \cdot\frac { \hat { \beta  } _{ 2 } }{ \omega _{ I } }  \right) \left( \frac { N_{ P }^{ * }\beta _{ 2 } }{ \delta  }  \right)  } .
\label{eq:rh_baseline}
\end{equation}

Rewriting the expression for $R_P$, we find that it is the average number of newly colonized patients as result of contacts with other colonized patients. As in the first factor of $R_H$, the expression for $R_P$ includes two modes of spread: direct spread characterized by colonized patient transmission, and indirect spread characterized by infected patient transmission. Thus,\\
\vspace{-1ex}
\begin{equation}
R_{ P }=N_{ P }^{ * }\left( \frac { \beta _{ 1 } }{ \alpha +\phi +\omega _{ C } } +\frac { \phi  }{ \alpha +\phi +\omega _{ C } } \cdot\frac { \beta _{ 3 } }{ \omega _{ I } }  \right) .
\end{equation}

\noindent
Furthermore, since $R_H, R_P>0$, we have from \eqref{eq:ro_form} that:
\begin{equation}
R_{ 0 }=\frac { R_{ P } }{ 2 } +\frac { 1 }{ 2 } \sqrt { R_{ P }^{ 2 }+4R_{ H }^{ 2 } } >\frac { R_{ P } }{ 2 } +\frac { 1 }{ 2 } \sqrt { R_{ P }^{ 2 } } =R_{ P }.
\label{eq:cond1}
\end{equation}

Applying the triangle inequality, we also find that:
\begin{equation}
R_0 = \frac{R_P}{2} + \frac{1}{2} \sqrt{R_P^2 + 4R_H^2} < \frac{R_P}{2} + \frac{1}{2}\left( R_P + 2R_H\right) = R_P + R_H \text{.}
\label{eq:cond2}
\end{equation}

Combining these results, we can say that, in general, $R_P < R_0 < R_P + R_H$. 
The latter part of this inequality means that the two infection potentials, $R_P$ and $R_H$, have a net effect (given by the adjusted reproduction number, $R_0$) which is less than their sum. This is explained by the fact that patients are capable of transmitting MRSA to both patients \textit{and} HCWs, while HCWs can only transmit MRSA to patients, and not to other HCWs. Recall that, as no new infected or colonized patients are being admitted into the system, this adjusted reproduction number accounts only for the spread of MRSA within  hospital facilities absent colonized/infected patient admission.

\subsubsection{Screening at Admission}

As before, we set $$\Lambda=\left( \gamma_{U} + \mu_{U} \right)U + \left( \gamma_{C} + \mu_{C} \right)C + \mu_{I} I + \left( \gamma_{Z} + \mu_{Z} \right)Z,$$ so as to achieve a constant in-hospital population. At the DFE we have

\begin{equation}
\begin{aligned}
{H_{C}}^{*}={C}^{*}={I}^{*}& ={Z}^{*} =0,\\
{H}^{*}&=N_{H},\\
{U}^{*}&=N_{P}.\\
\end{aligned}
\end{equation}
\noindent
Once again, the system can be simplified by making the substitutions $H = N_H - H_C$ and $U = N_P - U - C - I - Z$. For convenience, let $\Sigma = C + I + Z$, so that $U = N_P - \Sigma$. Using the next-generation matrix approach, we obtain:

\begin{equation*}
\mathscr{F}=\begin{pmatrix} \hat {\beta}_1 \frac { C(N_H - H_C)}{N} +\hat{\beta}_2 \frac { I(N_H-H_C)}{N}  \\[6pt] \beta_1 \frac {C(N_P - \Sigma)}{ N } + \beta_2 \frac{ H_C (N_P-\Sigma)}{N} + \beta_3 \frac{I(N_P - \Sigma)}{ N }  \\[6pt] 0 
\end{pmatrix}\\ 
\quad \text{and} \quad
\mathscr{V}=\begin{pmatrix} \delta { H }_{ C } \\[6pt] C(\omega_C + \phi + \alpha) \\[6pt] I(\mu_I+\kappa )-\phi C \\[6pt]
\end{pmatrix},
\end{equation*}
which yields
\begin{equation*}
F=\begin{pmatrix} 0 & \hat { { \beta  }_{ 1 } } { N }_{ H }^{ * } & \hat { { \beta  }_{ 2 } } { N }_{ H }^{ * } \\[6pt] { \beta  }_{ 2 }{ N }_{ P }^{ * } & { \beta  }_{ 1 }{ N }_{ P }^{ * } & { \beta  }_{ 3 }{ N }_{ P }^{ * }\\[6pt] 0 & 0 & 0\end{pmatrix}\\ 
\quad \text{and} \quad
V=\begin{pmatrix} \delta  & 0 & 0\\[6pt] 0 & \alpha + \omega_C +\phi  & 0\\[6pt] 0 & -\phi  & \kappa +{ \mu  }_{ I } \end{pmatrix}.
\end{equation*}
The next-generation matrix is then

\begin{equation}
F{ V }^{ -1 }=\begin{pmatrix} 0 & \frac { { N }_{ H }^{ * }\left[ \hat { { \beta  }_{ 1 } } \left( \kappa +{ \mu  }_{ I } \right) +\hat { { \beta  }_{ 2 } } \phi  \right]  }{ \left( \kappa +{ \mu  }_{ I } \right) \left( \alpha +{ \mu  }_{ C }+{ \gamma  }_{ C }+\phi  \right)  }  & \frac { { N }_{ H }^{ * }\hat { { \beta  }_{ 2 } }  }{ \kappa +{ \mu  }_{ I } }  \\[6pt] \frac { { N }_{ P }^{ * }{ \beta  }_{ 2 } }{ \delta  }  & \frac { { N }_{ P }^{ * }\left[ { \beta  }_{ 1 }\left( \kappa +{ \mu  }_{ I } \right) +{ \beta  }_{ 3 }\phi  \right]  }{ \left( \kappa +{ \mu  }_{ I } \right) \left( \alpha +{ \mu  }_{ C }+{ \gamma  }_{ C }+\phi  \right)  }  & \frac { { N }_{ P }^{ * }{ \beta  }_{ 3 } }{ \kappa +{ \mu  }_{ I } } \\[6pt] 0 & 0 & 0 \end{pmatrix}.
\end{equation}\\
The basic reproduction number has the same form as \eqref{eq:ro_form} and satisfies \eqref{eq:cond1} and \eqref{eq:cond2}. $R_P$ and $R_H$ are given by

\begin{equation}
{ R }_{ P }={ N }_{ P }^{ * }\left( \frac { { \beta  }_{ 1 } }{ \alpha +{ \mu  }_{ C }+{ \gamma  }_{ C }+\phi  } +\frac { \phi  }{ \alpha +{ \mu  }_{ C }+{ \gamma  }_{ C }+\phi  } \cdot\frac { { \beta  }_{ 3 } }{ \kappa +{ \mu  }_{ I } }  \right)
\end{equation} 
\text{ and }

\begin{equation}
{ R }_{ H }=\sqrt { { N }_{ H }^{ * }\left( \frac { \hat { { \beta  }_{ 1 } }  }{ \alpha +{ \mu  }_{ C }+{ \gamma  }_{ C }+\phi  } +\frac { \phi  }{ \alpha +{ \mu  }_{ C }+{ \gamma  }_{ C }+\phi  } \cdot\frac { \hat { { \beta  }_{ 2 } }  }{ \kappa +{ \mu  }_{ I } }  \right) \frac { { N }_{ P }^{ * }\\{ \beta  }_{ 2 } }{ \delta  }  },
\end{equation}
respectively. The difference between the baseline reproduction number and the reproduction numbers for the screening models is the introduction of the rate $\kappa$, which is the rate of progression of infected patients to the isolation unit. Note also that the discharge rate $\gamma_I$, which was implicit in the $\omega_I$ rate, is excluded.
A clear disadvantage of setting $\lambda_C=\lambda_I=0$ in the analysis is that the screening parameter $\rho$ does not appear in the expression for the adjusted reproduction number. The adjusted reproduction number here is the same as that found for the model with screening at discharge.

\subsection{Sensitivity analysis of \texorpdfstring{$R_0$}{Lg}}

\subsubsection{Baseline model}

\begin{figure}
    \centering
    \includegraphics[width=\textwidth]{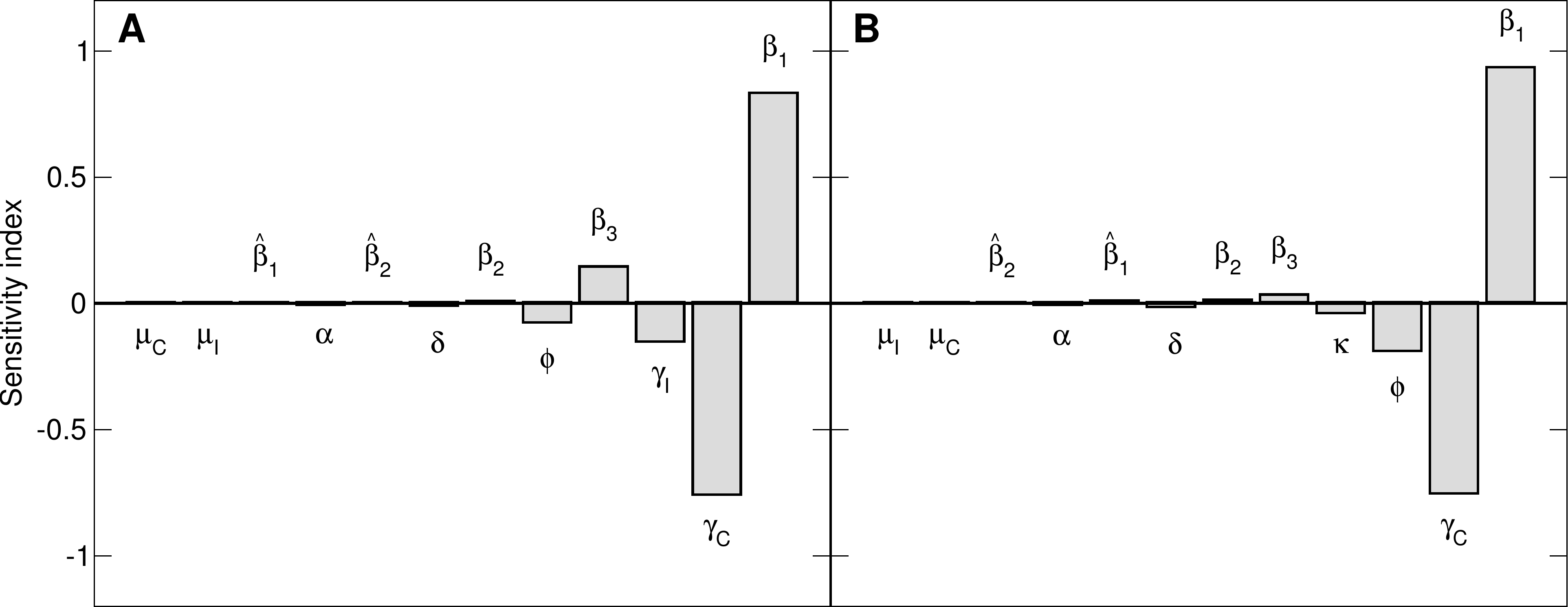}
    \caption{Sensitivity analysis of the adjusted reproduction number $R_0$ with respect to some parameters for (A) the baseline model and (B) the models with screening strategies.}
    \label{fig:sensitivity}
\end{figure}

For the sensitivity analysis, we assumed that the parameters were obtained from normal distributions. Local sensitivity indices are estimated from the partial derivatives of $R_0$. That is, a $1\%$ change in the specified parameter yields the given percent change in $R_0$ (see Figure \ref{fig:sensitivity}). 
Figure \ref{fig:sensitivity} summarizes the indices of sensitivity for the adjusted reproduction number of the baseline model, as it appears in equation \eqref{eq:r0_baseline}. 
As one can see, the parameters to which $R_0$ is most sensitive are the rate of transmission between uncolonized and colonized patients ($\beta_1$) and the discharge rate of colonized patients ($\gamma_C$).

In our baseline model there are several mechanisms by which an uncolonized patient may become colonized with MRSA. This sensitivity analysis proves that the most important mechanism to mitigate is direct/indirect colonization via other colonized patients. The parameter $\gamma_C$ summarizes the flow out from the colonized compartment due to treatment or discharge. This means that as more colonized patients leave the hospital, net MRSA transmission rate drops. In the baseline model, only these two parameters exhibited significant effects on $R_0$ when changed by small amounts. The remaining parameters produced negligible effects on $R_0$. 

\begin{figure}[!htb]
    \centering
    \includegraphics[width=0.6\textwidth]{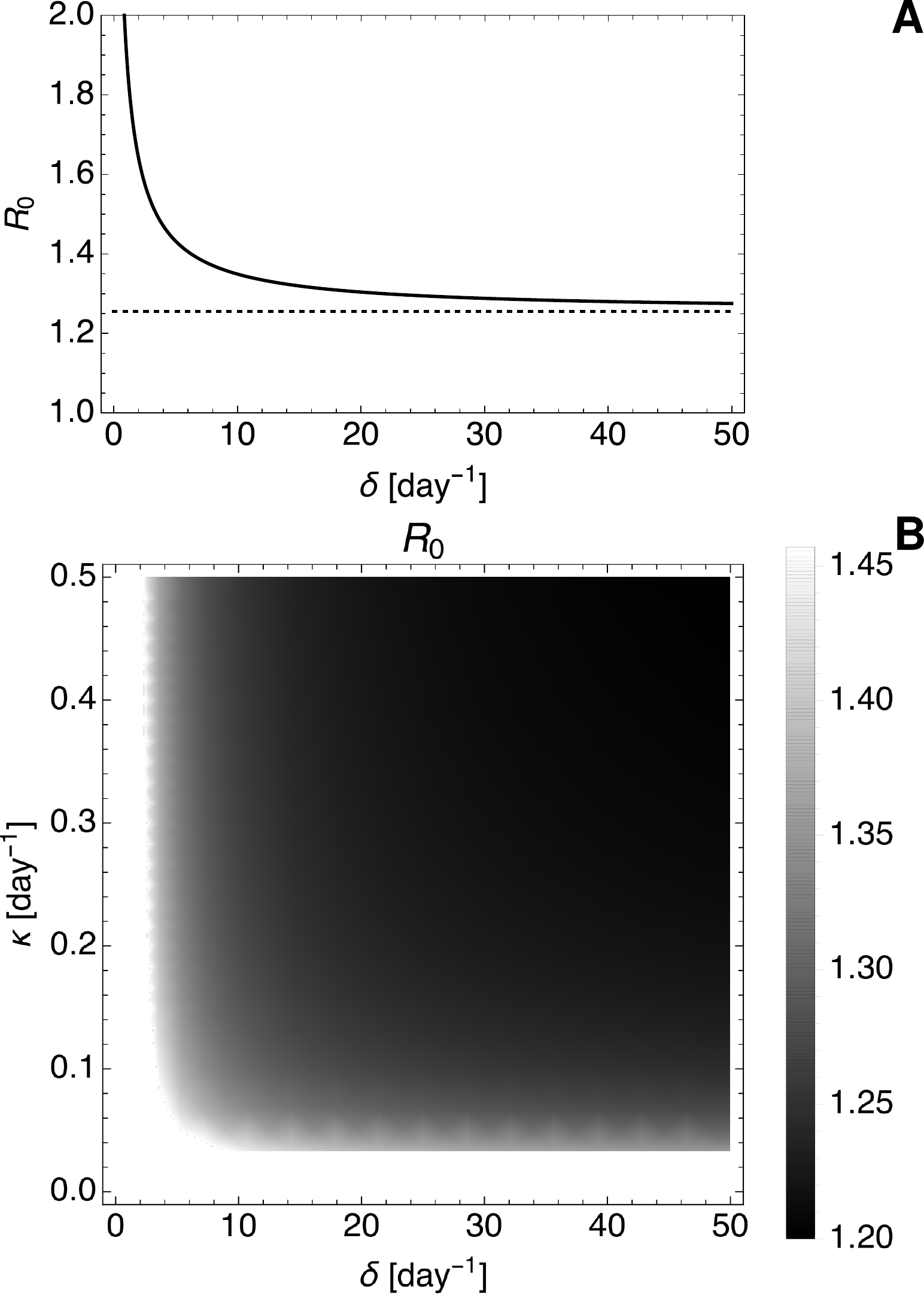}
    \caption{(A) The adjusted reproduction number $R_0$ of the baseline model is shown with respect to the decontamination rate of HCWs, $\delta$. The adjusted reproduction number is always bigger than $R_0\approx1.26$ (dotted line). (B) The adjusted reproduction number of the screening models is shown as a function of the decontamination rate of HCWs, $\delta$, and the rate at which infected patients move to isolation, $\kappa$. Note that $R_0 > 1$ for all values of $\delta$ and $\kappa$ plotted.}
    \label{fig:R0}
\end{figure}

The time it takes for a contaminated HCW to become decontaminated  can vary between $6$ hours and $24$ days \cite{grundmann2002risk}, and it can be seen in  Figure \ref{fig:R0}A that the adjusted reproduction number of the baseline model is always greater than $1$ for any of these values of $\delta$. This means that reducing the decontamination rate can decrease the value of the adjusted reproduction number, but it is never enough to prevent an outbreak in the absence of any other control effort. Small changes in $\delta$ produce negligible effects on outbreak likelihood, as shown by our sensitivity analysis. 


\subsubsection{Screening at admission and discharge}

For the models with screening at admission and discharge, the same sensitivity analysis was performed. Once again we found that $\beta_1$ and $\gamma_C$ were the most influential parameters (see Figure \ref{fig:sensitivity}B). A more careful analysis was performed for other parameters of interest. In the screening models, $1/\kappa$ denotes the average time it takes for an infected patient to move to isolation. To determine how both $\kappa$ and $\delta$ affected the reproduction number in the models with screening, the latter was graphed as a function of either variable. The results are shown in Figure \ref{fig:R0}B. We found that when reducing \textit{only} $\delta$ and $\kappa$, it is always the case that $R_{0}>1$. Therefore, in order to reduce the prevalence of MRSA in hospitals, it is not sufficient to only ensure that HCWs are adhering to hygiene policies and screening patients regularly for MRSA infection - it is necessary to control for other parameters.

Other model parameters were analyzed similarly (not shown). For $\delta > 5$ day$^{-1}$, $\beta_1 < 0.21$ day$^{-1}$ ensures that the reproduction number is less than $1$. If $\delta \le 5$ day$^{-1}$, however, this may not be sufficient in order to prevent endemicity of MRSA in the system. A similar phenomenon can be observed with the parameter $\gamma_C$. This indicates that as long as competent hygiene policies are enforced, an endemic free system can be achieved so long as $\beta_1, \gamma_C < 0.2$ day$^{-1}$ in a system with screening (i.e. strict precautions are taken when contacting colonized patients and colonized patients are treated as soon as possible, ideally at a rate of $1$ in every $5$ days). 


\subsection{Endemic Equilibria}

The endemic equilibrium corresponds to a steady state where the disease remains in the population \cite{brauer2008compartmental}. The high level of complexity associated with our model does not allow us to find a closed form solution for the endemic equilibria, hence results were found using numerical methods. In order to find the endemic equilibria of the baseline and screening models, we reduced the system of equilibrium conditions to one algebraic equation with one variable and to two algebraic equations with two variables, respectively. 

\begin{figure}[!htb]
    \centering
    \includegraphics[width=0.9\textwidth]{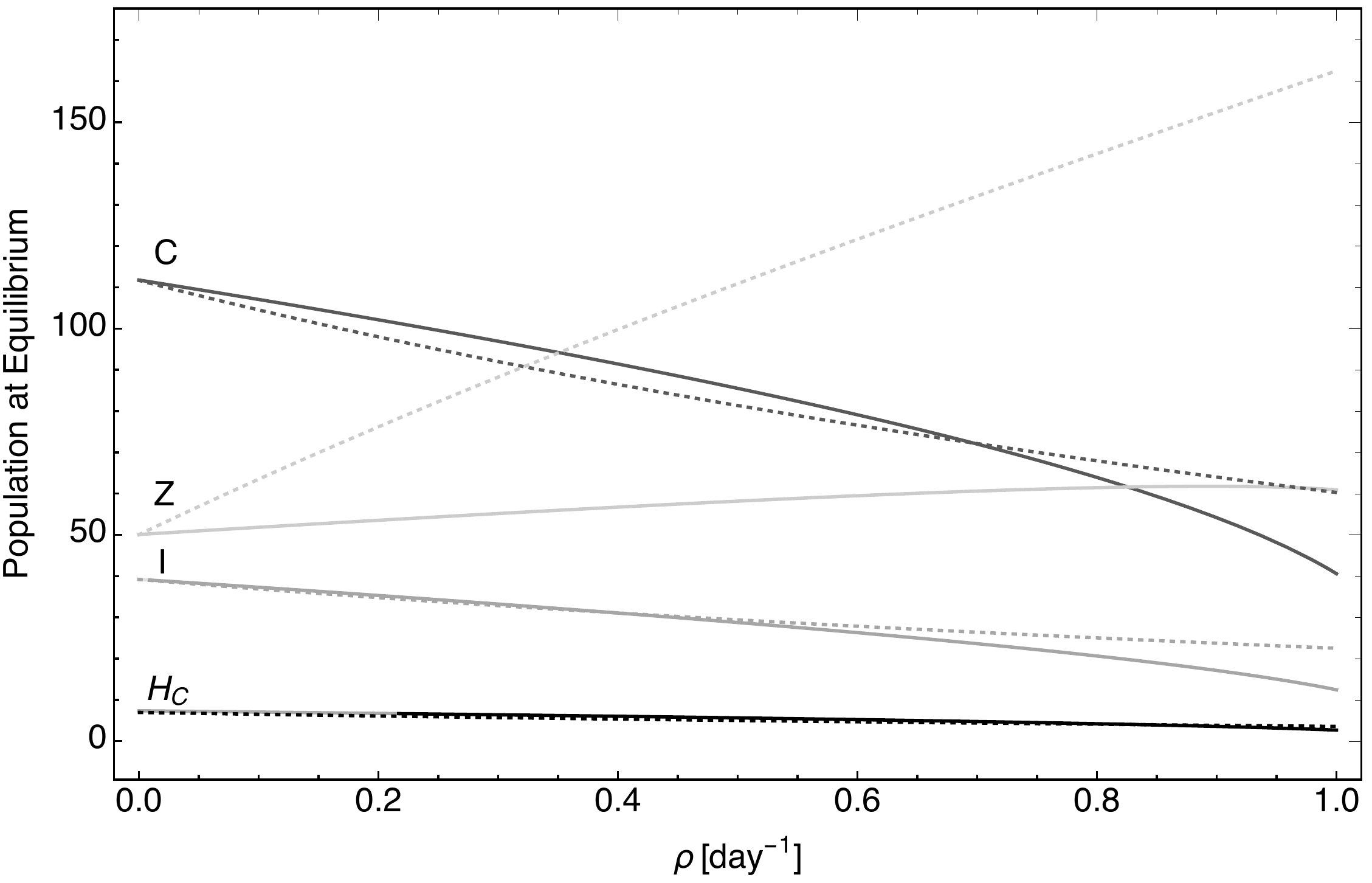}
    \caption{Population sizes at the endemic equilibrium as $\rho$ varies. The solid lines correspond to screening at admission, whereas the dotted lines correspond to screening at discharge.}
    \label{fig:endemic_eq}
\end{figure}

Figure \ref{fig:endemic_eq} shows the solutions to these resulting equations in terms of the screening proportion $\rho$. For screening proportions below $60\%$, screening at patient discharge produced marginally better endemic conditions than screening at patient admission. Although endemic conditions were more favorable for the model with screening at discharge, the isolated patient population was much larger than in the model with screening at admission. For higher screening proportions, screening at patient admission produced better endemic conditions. For all but the highest screening proportions ($\rho > 0.8$), contaminated HCW and infected patient populations were very similar in size for either screening model.

\begin{figure}[!htb]
    \centering
    \includegraphics[width=\textwidth]{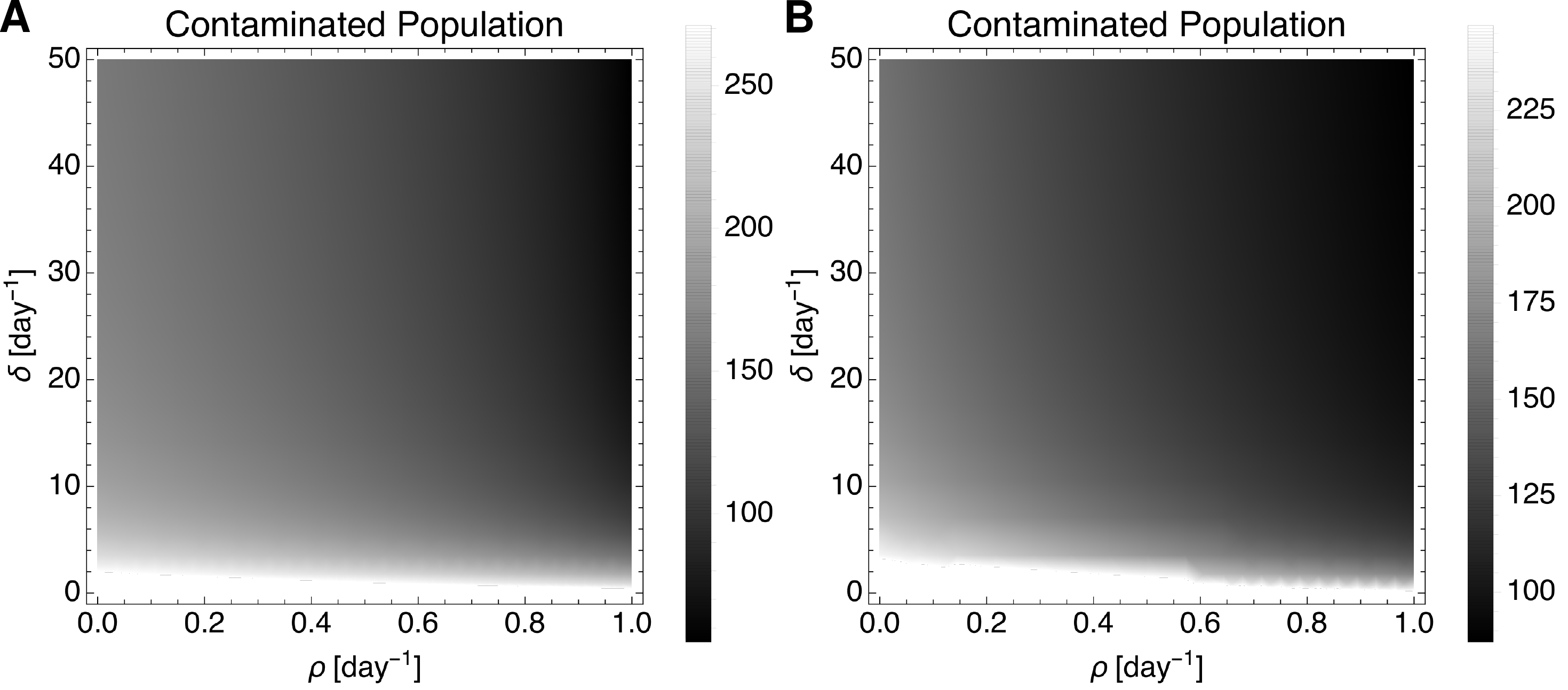}
    \caption{Contaminated patient population as a function of screening proportion ($\rho$) and the decolonization rate of contaminated HCWs ($\delta$) for the model with screening at (A) admission and (B) discharge.}
    \label{fig:endemic_eq_density}
\end{figure}

Figure \ref{fig:endemic_eq_density}A presents the sum of colonized patients, infected patients, and contaminated HCWs as a function of parameters $\rho$ and $\delta$ for the model with screening at admission. We shall refer to this sum as the contaminated population. Sufficiently large values of $\delta$ ($\delta > 10$ day$^{-1}$) produced little overall influence on the contaminated population when compared to $\rho$.

The same comparison was made for the model with screening at discharge in Figure \ref{fig:endemic_eq_density}B. There was little difference between screening models in how the parameter $\delta$ affected the contaminated population. Interestingly, even for very high screening proportions, the contaminated population in the model with discharge screening was still greater than $100$. This is a consequence of the mechanism of screening, as discharge screening does not stop individuals colonized or infected with MRSA from entering the hospital. For higher values of $\rho$ the model with screening at admission produced more favorable results, indicated by a smaller contaminated population.

\renewcommand{\thesubfigure}{\Alph{subfigure}}




\subsection{Results}

We developed and analyzed three models of nosocomial MRSA transmission to evaluate the superior screening strategy. Popular hospital practice favors screening at admission, while screening at discharge had yet to be tested. We modeled each system using a system of nonlinear differential equations. We found that, for screening values of $\rho < 0.6$ (i.e. less than $60\%$ of new/outgoing patients are screened at admission/discharge), screening at discharge produced marginally better results in the form of a smaller overall colonized patient population. For $\rho < 0.6$, the effect of either control strategy on the infected patient population was nearly equivalent. For $\rho > 0.6$, on the other hand, screening at admission was the better screening method, as it better controlled both colonized and infected patient populations. For all values $\rho > 0$, screening at discharge yielded more isolated patients than the alternative control strategy, suggesting cost limitations.

\section{Discussion}

MRSA prevalence in hospital facilities is a concern of increasing priority since it jeopardizes the health of patients and health care workers alike. However, MRSA cannot be treated exclusively with antibiotics due to the very realistic possibility of further resistant strains. Thus, control strategies and protocols should be emphasized in health care facilities so as to control bacterial spread and further proliferation. Screening followed by isolation is a very common method of controlling MRSA. Of practical consideration is the most effective means of screening. Here we evaluated the effectiveness of discharge screening as compared to the typical alternative of admission screening. In order to compare the two proposed strategies for MRSA control in hospitals, we evaluated three compartmental models: a baseline model and two models for either screening strategy.  The difference in the design of the models is intended to answer questions otherwise not addressed in the current literature regarding patients leaving hospitals and the effect on MRSA transmission dynamics in hospitals. 

Screening at discharge appears to be the more effective strategy in reducing endemic populations within a hospital for lower screening percentages. Although common practice prefers screening at admission, our results show that, for screening percentages below $60\%$, screening at discharge is more effective in reducing colonized patient populations within a hospital. For the same range of screening percentages, the infected patient and contaminated HCW populations were nigh indistinguishable (differences in equilibria were insignificant). However, discharge screening also yields a very rapid growth in the number of isolated patients, suggesting that the strategy may not be entirely practical if considering an IU with limited capacity. For screening percentages greater than $60\%$, screening at admission performed better overall, yielding lower equilibria for both the colonized and infected patient populations. Contaminated HCW populations were once again indistinguishable between models for higher values of $\rho$. We also made a continuous-time Markov chain version for each of the three models to test the effects of stochasticity, the results of which (not shown) mirrored those shown for the deterministic models. 

Some areas of further research and elaboration remain. The most significant of which includes an isolation unit (IU) with finite capacity (e.g. 20 beds). This consideration would clarify the practicality of discharge screening and resolve the issue of whether or not the growth in the number of isolated patients can be accommodated. Another important consideration is cost. Although we can mathematically express the results of the above models in a concise and simple manner, the true pragmatism must be evaluated in terms of cost. A significant problem associated with controlling MRSA is the cost it incurs in treatment and various methods to prevent its spread. These results might guide policy makers to improve control strategies, but a detailed cost analysis might produce more sound results. In performing our research, parameter values were chosen conservatively so as to provide a lower bound for any results later on. 

As mentioned earlier, we made the common assumption that patients and HCWs mix homogeneously, and that our patient and HCW populations were constant. To give a more detailed description of heterogeneous mixing would require significantly more data on HCW-patient contact patterns than is presently available. 

Parameters were taken, for the most part, from primary sources and various papers discussing MRSA endemic dynamics. A more exhaustive analysis could include confidence intervals and hypothesis testing. 
    
 \section{Acknowledgments}

We would like to thank Dr. Carlos Castillo-Chavez, Founder and Co-Director of the Mathematical and Theoretical Biology Institute (MTBI), for giving us the opportunity to participate in this research program. We would also like to thank Co-Director Dr. Anuj Mubayi as well as Coordinator Ms. Rebecca Perlin and Management Intern Ms. Sabrina Avila for their efforts in planning and executing the day-to-day activities while at MTBI. Finally, we would also like to thank Baltazar Espinoza and Victor Moreno for their help and advice. This research was conducted as part of 2018 MTBI at the Simon A. Levin Mathematical, Computational and Modeling Sciences Center (MCMSC) at Arizona State University (ASU). This project has been partially supported by grants from the National Science Foundation (NSF – Grant MPS-DMS-1263374 and NSF – Grant DMS-1757968), the National Security Agency (NSA – Grant H98230-J8-1-0005), the Alfred P. Sloan Foundation, the Office of the President of ASU, and the Office of the Provost of ASU.

\bibliographystyle{plain}
\bibliography{biblio}

\end{document}